\documentclass[aps,graphicx,twocolumn]{revtex4}

\usepackage{graphicx}
\usepackage{epstopdf}
\usepackage{verbatim}
\usepackage{afterpage}
\usepackage{amsmath}
\usepackage{float}
\usepackage[colorlinks,linkcolor=blue,citecolor=blue,urlcolor=blue]{hyperref}
\usepackage[caption=false,font=scriptsize,labelfont=sf,textfont=sf]{subfig}

\begin{document}
\title{Receiver-device-independent quantum secure direct communication}

\author{Cheng Liu$^{1}$, Cheng Zhang$^{2}$, Shi-Pu Gu$^{2}$, Xing-Fu Wang$^{1}$, Lan Zhou$^{1}$\footnote{Email address: zhoul@njupt.edu.cn}, Yu-Bo Sheng$^{2}$}
\address{
$^1$College of Science, Nanjing University of Posts and Telecommunications, Nanjing, 210023, China\\
$^2$College of Electronic and Optical Engineering, \& College of Flexible Electronics (Future Technology), Nanjing University of Posts and Telecommunications, Nanjing, 210023, China\\
}
\date{\today}

\begin{abstract}
Quantum secure direct communication (QSDC) enables the message sender to directly send secure messages to the receiver through the quantum channel without keys. Device-independent (DI) and measurement-device-independent (MDI) QSDC protocols can enhance QSDC's practical security in theory. DI QSDC requires extremely high global detection efficiency and has quite low secure communication distance. DI and MDI QSDC both require high-quality entanglement. Current entanglement sources prepare entangled photon pairs with low efficiency, largely reducing their practical communication efficiency. In the paper, we propose a single-photon-based receiver-device-independent (RDI) QSDC protocol. It only relies on the trusted single-photon source, which is nearly on-demand under current technology, and treats all the receiving devices in both communication parties as ``black-boxes''. The parties ensure the message security only from the observed statistics. We develop a numerical method to simulate its performance in practical noisy communication situation. RDI QSDC provides the same security level as MDI QSDC. Compared with DI and MDI QSDC, RDI QSDC has some advantages. First,
it uses the single-photon source and single-photon measurement, which makes it obtain the practical communication efficiency about 3415 times of that in DI QSDC and easy to implement. The whole protocol is feasible with current technology. Second, it has higher photon loss robustness and noise tolerance than DI QSDC, which enables it to have a secure communication distance about 26 times of that in DI QSDC. Based on above features, the RDI QSDC protocol makes it possible to achieve highly-secure and high-efficient QSDC in the near future.
\end{abstract}
\maketitle

\section{Introduction}\label{Section1}
Quantum communication aims to realize the secure message transmission based on the fundamental principles of quantum mechanics. Quantum communication has the ability to detect the eavesdropping, and thus has unconditional security in theory. Quantum communication began with the researches on quantum key distribution (QKD) \cite{QKD1,QKD2,QKD3,QKD4,QKD5,QKD5n,QKD6,QKD7,QKD8}, which is used to distribute random secure keys between two remote communication parties. Besides QKD, quantum secure direct communication (QSDC) is another quantum communication mode \cite{QSDC1,QSDC2,QSDC3}. QSDC enables the message sender to directly transmit secret messages to the receiver through quantum channels without sharing keys in advance.

 QSDC was first proposed in 2000, exploiting the quantum entanglement and block transmission technique \cite{QSDC1}. Later, the entanglement-based QSDC protocol and the single-photon-based QSDC protocol were successively proposed, elucidating the conditions and physical mechanisms of QSDC \cite{QSDC2,QSDC3}. Since then, QSDC has achieved great progress in theory and experiment \cite{wang,capacity1,li,DIQSDC1,DIQSDC2,DIQSDC3,MDIQSDC1,MDIQSDC2,MDIQSDC3,masking,cao,one-step-QSDC,wu,zeng,zhang,pan,liu,wei,ying,QSDCe1,QSDCe2,QSDCe2n,QSDCe3,QSDCe4,QSDCe5,QSDCe6,QSDCe7,QSDCe8}. In theory, high-dimension QSDC \cite{wang}, one-step QSDC \cite{one-step-QSDC} and passive decoy-state QSDC \cite{ying} protocols are successively proposed. In the experimental aspect, the single-photon based QSDC and entanglement-based QSDC were experimentally demonstrated in 2016 and 2017, respectively \cite{QSDCe1,QSDCe2}. Later, the free-space QSDC \cite{QSDCe3}, QSDC network \cite{QSDCe4}, 100 kilometers QSDC \cite{QSDCe5}, and continuous-variable QSDC \cite{QSDCe7,QSDCe8} experiments have been successively reported. Recently, the quantum networks with secure classical repeaters adopted QSDC to realize the secure end-to-end communication across the entire network \cite{QSDCe6}.

QSDC has the unconditional security in theory. However, imperfect devices provide the eavesdropper (Eve) the opportunity for stealing messages without being detected. This motivates the investigation of the stronger, device-independent (DI) QSDC \cite{DIQSDC1,DIQSDC3,zeng,DIQSDC2}. DI QSDC treats all devices in each party as a ``black-box'', and guarantees the communication security relying only on the observation of nonlocal quantum correlations. DI QSDC can resist all possible attacks on imperfect devices, and provides QSDC the highest security. However, similar as the other DI protocols \cite{DIQKD1,DIQKD2,DIQKD3,DIQKD4,DIQSS}, DI QSDC relies on the distribution of high-quality entanglement. It requires extremely high global detection efficiency (about 0.926) and has low noise tolerance, which greatly increases its experimental difficulty and limits its secure communication distance (less than 1 km) \cite{DIQSDC1}. Measurement-device-independent (MDI) QSDC hands all the measurement tasks to an untrusted measurement party \cite{MDIQSDC1,MDIQSDC2}. MDI QSDC is immune to all possible attacks from the imperfect measurement devices. Combined with the decoy-state method, MDI QSDC can guarantee the practical security of QSDC in theory. However, MDI QSDC also requires high-quality entanglement source and Bell-state measurement (BSM). The
entanglement generation of practical entanglement source
(spontaneous parametric down-conversion source) is low efficient ($10^{-5}-10^{-3}$ \cite{Hu}). Meanwhile, the success probability of BSM with linear optical element is only 50\%. These features reduce DI QSDC's and MDI QSDC's efficiency and limit its practical implementation.

In 2022, the receiver-device-independent (RDI) QKD was proposed, where the key sender's
device is (partially) trusted while the receiver's
device  can be treated as a ``black-box'' \cite{RDIQKD1}. The RDI QKD achieves the one-sided DI security and is amenable to a
practical prepare-and-measure implementation. RDI QKD achieved the proof-of-principle demonstration soon \cite{RDIQKD2}. In the paper, inspired by the RDI QKD, we propose the single-photon-based RDI QSDC protocol. As the practical single-photon sources have already allowed for nearly on-demand \cite{demand},
highly-efficient \cite{highlyefficient} extraction of single photons
  (including in pulse sequences \cite{solid1,solid2} and at telecommunications frequencies \cite{frequencies}), and can maintain the purity and indistinguishability of the emitted photons above 99\% \cite{99,ondemand}, we assume that the single-photon source is trusted. We treat all the receiving devices (including the quantum memory (QM) and measurement devices) in both communication parties as ``black-boxes''. The communication parties can continuously verify (or self-test) the correct
operation of all the devices and bound the message leakage rate to Eve only using the classical inputs (basis selections) and outputs (measurement results) of the receiving devices, similarly to the full DI model. We develop a numerical method to simulate its performance in practical noisy environment. The RDI QSDC can achieve the same security level as MDI QSDC.
Compared with DI and MDI QSDC, RDI QSDC has some advantages. First, it only uses the single-photon source and single-photon measurement (SPM), which makes it have much higher practical communication efficiency and easy to implement. The whole RDI QSDC protocol is feasible with current technology. Second, RDI QSDC has higher photon loss robustness and noise tolerance than DI QSDC, which enable it to have much longer secure communication distance. It can achieve a secure communication distance about 26 times of that in DI QSDC.
 Based on above features, our RDI QSDC protocol makes it possible to achieve highly-secure and high-efficient QSDC in the near future.

The paper is organized as follows. In Sec. \ref{Section2}, we provide the basic principle of the RDI QSDC protocol and its implementation with linear optical elements. In Sec. \ref{Section3}, we show that the RDI QSDC protocol can resist the blinding attack. In Sec. \ref{Section4}, we develop a numerical method to estimate RDI QSDC protocol's performance in practical noisy environment. Finally, we make some discussion and draw a conclusion in Sec. \ref{Section5}.

\section{RDI QSDC protocol}\label{Section2}
 \subsection{The theoretical RDI QSDC protocol}\label{Section2.1}
\begin{figure*}[t]
\centering
\includegraphics[scale=0.36]{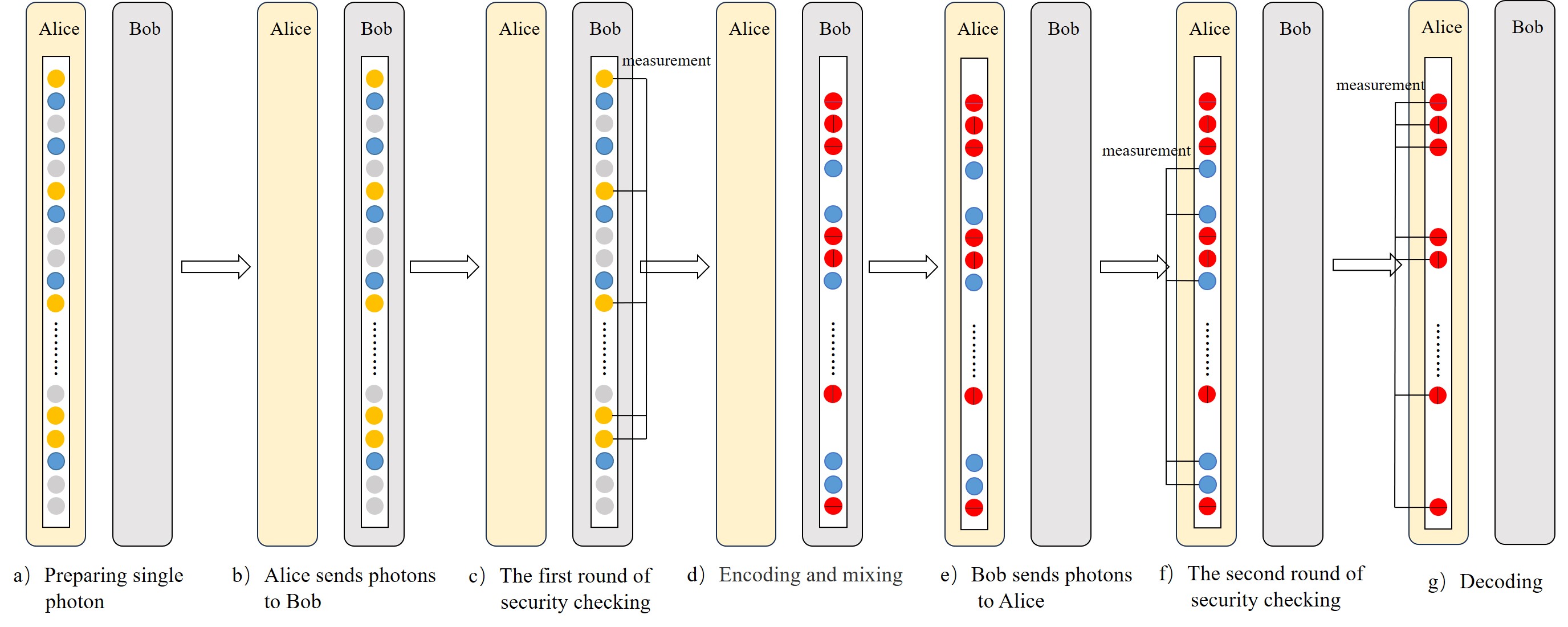}
\caption{Schematic diagram of the RDI QSDC protocol. Here, the orange, blue, and gray circles represent the photons in sequence $S_1$, $S_2$ and $S_3$, respectively. The circles connected by solid lines represent the photons being measured at this stage. The red circles with the horizontal and vertical lines represent the encoded photons with $U_0$ and $U_1$ operations, respectively. }
\label{fig:boat2}
\end{figure*}

The RDI QSDC protocol is based on only three basic security assumptions. (i) The quantum mechanics is correct. (ii) no unwanted measurement results or encoding operations can escape from the communication
parties' laboratories. (iii) the photon source is trusted. Besides, we do not rely on any physical model of the receiving devices. The RDI QSDC protocol includes the following 6 steps.

\emph{\textbf{Step 1}}
The message receiver Alice has an integer sequence $\left\{X\right\}=\left\{x_1,x_2,\cdots,x_n \right\}$ containing $n$ positive integers $x_i=i$ $\left(i\in\{ 1, 2,\cdots,n\}\right)$. Here, it is required that $n\geq3$ and $n\neq 4$ (The reason for $n\neq 4$ is shown in the discussion.)
Each positive integer $x_i$ can be used to generate a single photon state $|\psi_{x_i}\rangle=\cos\theta|0\rangle+e^{\frac{i2\pi{x_i}}{n}}\sin\theta|1\rangle$. Here, $|0\rangle$ and $|1\rangle$ represent two orthogonal quantum states, and we set $\theta=\frac{\pi}{4}$. Alice randomly selects the integer $x_i$ $\left(i\in\{1, 2,\cdots,n\}\right)$ to prepare $3r$ single photons ($r$ is a large number). She randomly selects $r$ photons to form the photon sequence  $S_1$ for the first round of security detection, and $r$ photons to form photon sequence $S_2$ for the second round of security detection. The rest $r$ photons form the message transmission photon sequence $S_3$. The integer sequences corresponding to the photon sequences $S_1$, $S_2$ and $S_3$ are denoted as $\{X_1\},\{X_2\},\{X_3\}$, respectively, which can be described as  $\{X_1\}=\{a_1,a_2,a_3,\cdots,a_r\}$, $\{X_2\}=\{b_1,b_2,b_3,\cdots,b_r\}$, $\{X_3\}=\{c_1,c_2,c_3,\cdots,c_r\}$.

Then, Alice performs random operation $U_0$ or $U_1$ on the photons in sequence $S_3$, where
\begin{eqnarray}\label{caozuo}
	U_0=I=\begin{bmatrix}
		1&0\\
		0&1\\
	\end{bmatrix},U_1=\sigma_z=\begin{bmatrix}
	1&0\\
	0&-1\\
	\end{bmatrix}.
\end{eqnarray}
$U_0$ and $U_1$ represent identity operation and phase-flip operation, respectively.

\emph{\textbf{Step 2}}
Alice sends all the photons to the information sender  Bob successively  through a quantum channel. After receiving the photons, Bob stores them in the quantum memory (QM).

 \emph{\textbf{Step 3}}
 Alice announces the positions of the photons in sequence $S_1$. Bob extracts these photons  from the QM. Then, the parties perform the first round of security checking (including the security of the photon transmission and the correct operations of the devices). In detail, Bob constructs an integer sequence $\left\{Y\right\}= \left\{ y_1, y_2,\cdots, y_n \right\}$, where $y_i=i$ $\left(i\in\{ 1, 2,\cdots,n\}\right)$. Bob randomly selects integers in $\left\{Y\right\}$ to  form the integer sequence $\{Y_1\}=\{w_1,w_2,w_3,\cdots,w_r\}$, where $w_m\in \{Y\}$ ($m=1,2,\cdots,r$). For the $mth$ security checking photon, Bob uses the integer $w_m$ to perform the measurement $M_{w_m}=|\psi_{w_m}\rangle\langle\psi_{w_m}|$, where $|\psi_{w_m}\rangle=\cos\theta|0\rangle+e^{\frac{i2\pi{w_m}}{n}}\sin\theta|1\rangle$. In this way, $M_{w_m}$ has the form of
\begin{eqnarray}\label{celiang1}
M_{w_m}&=&|\psi_{w_m}\rangle\langle\psi_{w_m}|\nonumber\\
&=&\cos^2\theta|0\rangle\langle0|+e^{\frac{-i2\pi{w_m}}{n}}\sin\theta\cos\theta|0\rangle\langle1|\nonumber\\		
&+&e^{\frac{i2\pi{w_m}}{n}}\sin\theta\cos\theta|1\rangle\langle0|+\sin^2\theta|1\rangle\langle1|.
\end{eqnarray}
We assume each measurement has binary outputs $g=0,1$. The output $g=0$ indicates that the measured quantum state is projected into $|\psi_{w_m}\rangle$, while the output $g=1$ indicates that the measured quantum state is projected into $I-|\psi_{w_m}\rangle$.

After the measurements for all the security checking photons,
 Bob announces the sequence $\{Y_1\}$ and the measurement output for each security checking photon. Alice calculates the theoretical probability distribution of the outputs based on the preparation basis and measurement basis of each photon as
 \begin{eqnarray}\label{gailv1}
 	\begin{aligned}
 	\begin{split}
 	&P_1(g=0)=\frac{\sum_{i=1}^{r}|\cos^2\theta+e^{\frac{i2\pi(a_i-w_i)}{n}}\sin^2\theta|^2}{r},\\
 	&P_1(g=1)=\frac{r-\sum_{i=1}^{r}|\cos^2\theta+e^{\frac{i2\pi(a_i-w_i)}{n}}\sin^2\theta|^2}{r}.
 \end{split}
\end{aligned}
 \end{eqnarray}

 She then compares this theoretical distribution with the practical probability distribution of measurement outputs. If the deviation between the practical probability distribution  and theoretical distribution exceeds the tolerable threshold, the communication has to be terminated. However, if the deviation is within the tolerable threshold, the communication proceeds to the next step.

\emph{\textbf{Step 4}}
Alice announces the positions of the photons in sequence $S_3$, and Bob extracts the photons from the QM. He encodes the messages on the extracted photons sequentially by performing the unitary operations $U_0$ and $U_1$ on the photons. Here, $U_0$ represents the classical message ``$0$'' and  $U_1$ represents the classical message ``$1$''.

After encoding, Bob extracts the photons from $S_2$ sequence from the QM, mixes the photons in $S_2$ and $S_3$ sequences to form $S_2'$ and $S_3'$ sequences, and then sends all the photons back to Alice successively through the quantum channel. The preparation basis sequence of the photons in $S_2'$ sequence is provided by
 \begin{eqnarray}\label{xulie}
	\{X_4\}=\{d_1,d_2,d_3,\cdots,d_r\}.
\end{eqnarray}

\emph{\textbf{Step 5}}
Alice receives all the photons and stores them in the QM. Then, Bob announces the positions of the photons in sequence $S_2'$. Alice extracts the corresponding photons from the QM and measures them based on her original preparation basis sequence $\{X_2\}=\{b_1,b_2,b_3,\cdots,b_r\}$. In this way, Alice's measurements are generated with the form of
\begin{eqnarray}\label{celiang2}
	\begin{aligned}
		\begin{split}
		M^{b_i}&=|\psi_{b_i}\rangle\langle\psi_{b_i}|\\
&=\cos^2\theta|0\rangle\langle0|+e^{\frac{-i2\pi{b_i}}{n}}\sin\theta\cos\theta|0\rangle\langle1|\\		
		&+e^{\frac{i2\pi{b_i}}{n}}\sin\theta\cos\theta|1\rangle\langle0|+\sin^2\theta|1\rangle\langle1|.\\
		\end{split}
	\end{aligned}
\end{eqnarray}
After all the measurements, Bob announces the original order of the photons in sequence $S_2$. If the measured quantum state is projected into $|\psi_{b_i}\rangle$, the measurement result is denoted as $g=0$. If the measured quantum state is projected into $I-|\psi_{b_i}\rangle$, Alice denotes the output $g=1$. Then, Alice calculates the theoretical probability distribution and practical probability distribution of the measurement results based on the preparation basis and measurement basis of each photon as
\begin{eqnarray}\label{gailv2}
	\begin{aligned}
		\begin{split}
			&P_2(g=0)=\frac{\sum_{i=1}^{r}|\cos^2\theta+e^{\frac{i2\pi(d_i-b_i)}{n}}\sin^2\theta|^2}{r},\\
			&P_2(g=1)=\frac{r-\sum_{i=1}^{r}|\cos^2\theta+e^{\frac{i2\pi(d_i-b_i)}{n}}\sin^2\theta|^2}{r}.
		\end{split}
	\end{aligned}
\end{eqnarray}

Then, Alice compares the theoretical distribution probabilities with the practical probability distribution probabilities of her measurement outputs. If the derivation exceeds the tolerable threshold, the communication should be terminated. If the deviation is within the tolerable threshold, the communication proceeds to the next step.

\emph{\textbf{Step 6}}
Bob announces the original positions of the photons in sequence $S_3'$. Alice extracts all the encoded photons and recovers the original message transmission photon sequence $S_3$. Then, she measures each photon with its original preparation basis. By comparing the photon state with its initial state, Alice can decode the message transmitted from Bob.

\subsection{RDI-QSDC implementation with linear optical elements}
\begin{figure*}[t]
\centering
\includegraphics[scale=0.17]{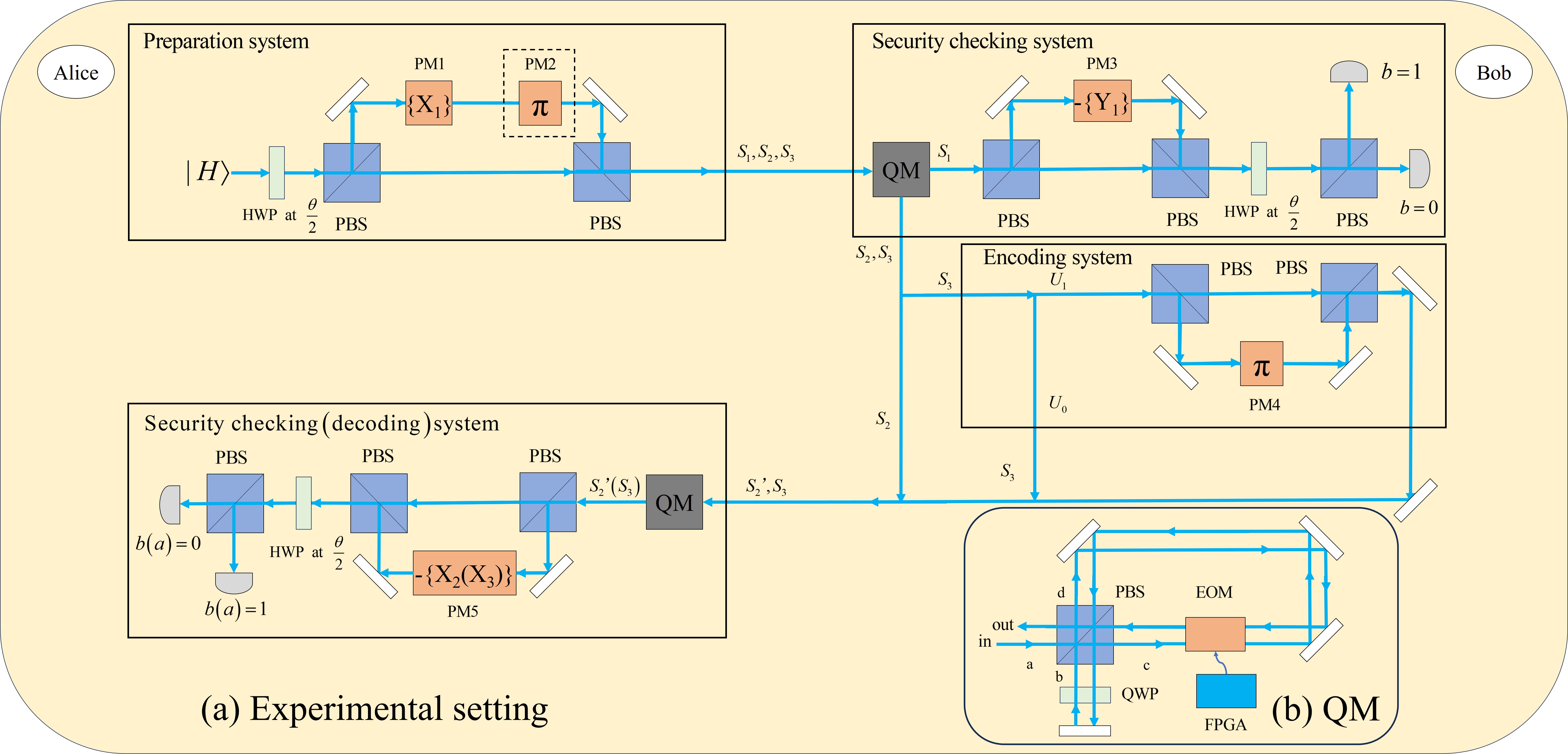}
\caption{(a) The implementation of the RDI QSDC protocol with linear optical elements. We use the polarization degree of freedom of photons to encode the messages. The half wave plate (HWP) rotated at $\frac{\theta}{2}$ can transform $|H\rangle$ into $\cos\theta|H\rangle+\sin\theta|V\rangle$. The polarization beam splitter (PBS) can totally transmit the photon in $|H\rangle$ and totally reflect the photon in $|V\rangle$. The phase modulators PM1, PM3, and PM5 are used to realize the preparation basis and measurement basis selection. PM2 and PM4 are used to perform the encoding. PM2 enclosed by dashed lines is used only when the unitary operation is $U_{1}$. (b) The structure of the all-optical polarization sensitive storage loop quantum memory (QM). The on-off of the electro-optic modulator (EOM) can control the storage and read out of the photons. The field programmable gate array
(FPGA) is used to control the on-off of the EOM. QWP represents the quarter-wave plate. By passing the photon through QWP twice, the polarization feature of the photon can be flipped.}
\label{Fig2}
\end{figure*}

The RDI QSDC protocol can be implemented with practical linear optical elements. As shown in Fig. \ref{Fig2}(a), the polarization degree of freedom of photons is used. $|H\rangle$ (horizontal polarization) represents $|0\rangle$ and $|V\rangle$ (vertical polarization) represents $|1\rangle$. Alice has two systems, say, the preparation system and security checking (decoding) system. First, Alice prepares photons in the quantum state of $|H\rangle$, which can be transformed into  $\cos\theta|H\rangle+\sin\theta|V\rangle$ after passing through the half-wave plate (HWP) set at $\frac{\theta}{2}$ (Here, we control $\theta=\frac{\pi}{4}$). Then, each photon successively passes through two polarization beam splitters (PBSs), which totally transmit the photon in $|H\rangle$ and totally reflect the photon in $|V\rangle$. By adjusting the phase modulator PM1 in the reflection path, Alice can prepare the quantum state $|\psi_{x_i}\rangle=\cos\theta|H\rangle+e^{\frac{i2\pi{x_i}}{n}}\sin\theta|V\rangle$. In addition, for each photon in sequence $S_3$, Alice randomly uses PM2 to shift its phase of $\pi$.

The photons in $S_1$, $S_2$ and $S_3$ sequences are sent to Bob through the quantum channel. Bob stores the received photons in the QM and extracts the corresponding photon sequences based on Alice's announcement. Bob has two systems, say, the security checking system and encoding system. The photons in sequence $S_1$ enter the security checking system. The basis selection process is controlled by PM3. Then, Bob passes the photon through the HWP set at $\frac{\theta}{2}$ and measures the output photon in two orthogonal polarization states for the security checking. The photons in sequence $S_3$ enter the encoding system. If the message is 0, Bob does not perform any operation on the corresponding photon. If the message is 1, Bob introduces the phase shift of $\pi$ with PM4 on $|V\rangle$ component to realize the phase-flip operation. For the photons in sequence $S_2$ and $S_3$, Bob mixes their orders to form new sequences $S_2'$ and $S_3'$. All the photons in $S_{2}'$ and $S_3'$ sequences are sent back through quantum channel to Alice.

 The security checking system and decoding system in Alice's location use the same set of devices. The photons in $S_2'$ sequence enter Alice's security checking system. Alice uses PM5 to realize the measurement basis selection for the photons in $S_{2}'$. Once the security checking is passed, the photons in $S_{3}$ sequence enter the decoding system. Alice uses PM5 to control the measurement basis for the encoded photons and finally reads out the messages.

 The QM adopted in our protocol is constructed with the all-optical, polarization insensitive storage loop with the structure in Fig. \ref{Fig2}(b) \cite{QM}. By controlling the on-off of the electro-optic modulator (EOM), the parties can make the entering photon circulate in the storage loop to achieve the  photon
storage, and read out the photon from the storage loop (details of the storage and readout processes are shown in Appendix A). This all-optical storage loop QM is feasible with current experimental technology.

\section{The security of the RDI QSDC protocol against blinding attack}\label{Section3}
 Our RDI QSDC protocol eliminates all additional assumptions on both parties' receiving devices (quantum memory and measurement devices). The parties only depend on the probability distribution of the measurement results (classical outputs) under various measurement bases (classical inputs) to assure the communication security. Our RDI QSDC protocol can resist all possible attacks from imperfect receiving devices, so that it provides the same security level as MDI QSDC. Here, we analyze its security against the most common blinding attack.

In the case of the blinding attack, Eve injects continuous light to each of Alice's and Bob's photon detectors. The injected light generates a huge photocurrent through each detector, which lowers the bias voltage and pulls each photon detector back to the linear mode. In the linear mode, the detector is insensitive to a single photon, but only sensitive to the power of the input pulse \cite{blind}. Then, Eve conducts a fake-state attack as follows. In the first photon transmission process, Eve intercepts some photons and randomly selects measurement basis to measure them. Then, Eve generates trigger pulses based on his measurement result with the power being slightly greater than the detector's response threshold. He sends the generated pulses to Bob. After Bob's encoding, Eve only needs to intercept an encoded photon from each pulse during the second round of photon transmission, and uses the preparation basis to measure it. By comparing the measurement result with the initial state, he can finally obtain the encoded message.

In conventional QSDC protocol, i.e., DL04 protocol, the blinding attack combined with fake-state attack does not increase the error rate of security checking \cite{blind}, so that it can not detected by the parties. However, in our RDI QSDC protocol, Eve's eavesdropping behavior can be detected by Alice and Bob during the security checking. In detail, the power of the pulse generated by Eve is slightly higher than the detector's response threshold. In Bob's location, each measurement basis corresponds to two detectors, which can detect the mutually orthogonal quantum states.
Bob's detectors will deterministically click $g=0$ when Bob's measurement basis is quite close to Eve's measurement basis, or deterministically click $g=1$ when Bob's measurement basis is nearly orthogonal to Eve's measurement basis.

We assume that Eve launches the blinding attack combined with the fake-state attack with probability of $p_1$ and chooses a measurement basis similar to Bob's basis with the probability of $p_1p_2$. We take the first round of security checking process as an example, the practical probability distribution of $g=0$ changes to
 \begin{eqnarray}\label{blinding attack}
	\begin{aligned}
		\begin{split}
		\frac{\sum_{i=1}^{(1-p_1)r}|\cos(\theta)^2+e^{\frac{i2\pi(a_i-w_i)}{n}}\sin(\theta)^2|^2}{2r}+\frac{1}{2}p_1p_2.
		\end{split}
	\end{aligned}
\end{eqnarray}
From Eq. (\ref{blinding attack}), the blinding attack combined with fake-state attack would change the practical probability distribution of $g=0$. With the growth of $p_1p_2$, the actual probability distribution of $g=0$ will tend toward 50\%. In this way, the more the theoretical probability distribution $P_1(g=0)$ deviates by 50\%, it is easier for the parties to detect the eavesdropping. With the growth of the basis number $n$, there will be more basis combinations for Alice, Bob, and Eve, so that the security checking will be more sensitive for Eve's attack. As a result, our RDI QSDC protocol can effectively resist the blinding attack combined with the fake-state attack.

\section{The numerical simulation of RDI QSDC's secrecy message capacity in practical noisy environment }\label{Section4}
Here, we consider the ideal single photon source and evaluate the performance of our RDI QSDC protocol under practical noisy condition, we define its secrecy message capacity $C_S$ as the ratio of the correctly and securely transmitted message bits to the total number of encoded quantum bits. Under ideal experimental condition, single photons can reach the communication parties perfectly during each photon transmission process. Each single photon can transmit 1 bit of message from Bob to Alice, so that $C_S=1$ and the secure communication distance is infinite. However, the practical noisy quantum channel  may cause photon loss and quantum state error, which will reduce $C_S$ and the secure communication distance.

According to the Wyner's wiretap channel theory, RDI QSDC's $C_S$ under a general collective attack can be calculated as \cite{Wyner,capacity1,QSDCe3,wei}
\begin{eqnarray}\label{gailv2}
		C_S=I(A:B)-I(B:E),
\end{eqnarray}
where $I(A:B)$ represents the mutual information between Bob and Alice, and $I(B:E)$ represents the mutual information between the message sender Bob and Eve.

 $I(A:B)$ can be calculated as \cite{QSDCe3,wei}
\begin{eqnarray}\label{I(B:A)}
	I(B:A)=Q_{ABA}[1-h(E_{ABA})],
\end{eqnarray}
 where $Q_{ABA}$ is the total gain of the detectors at Alice's location caused by the photons sent back from Bob, and $E_{ABA}$ represents the total error rate after two rounds of photon transmission. $h(x)=-x\log_2x-(1-x)\log_2(1-x)$ is the binary Shannon entropy.

Considering the characteristics of QSDC, Eve can obtain the transmitted  messages only if he can intercept corresponding photons in both photon transmission process. Therefore, we use the gain $Q_{AB}$ in Bob's location and the total error rate $E_{AB}$ after the first photon transmission to bound $I(B:E)$ as
\begin{eqnarray}\label{I_BE}
	\begin{aligned}
		\begin{split}
			I(B:E)\leq Q_{AB}*h(E_{AB}).
		\end{split}
	\end{aligned}
\end{eqnarray}

We denote the coupling coefficience between the photon and the optical fiber as $\eta_c$, the transmission efficiency of the photon in the channel as $\eta_t=10^{-\alpha L/10}$ ($\alpha=0.2 dB/km$), the storage efficiency of the QM as $\eta_m$, and the detection efficiency of the photon detector as $\eta_d$.
   $Q_{AB}$ and $Q_{ABA}$ can be described as
 \begin{eqnarray}\label{tance}
	Q_{AB}&=&\eta_t\eta_c\eta_m\eta_d,\nonumber\\
	Q_{ABA}&=&\eta_t^2\eta_c^2\eta_m^2\eta_d.
\end{eqnarray}

In both security checking processes, we can calculate the total error rates $E_{AB}$ and $E_{ABA}$ in two steps. In step one, we consider the photon does not lose, say, the party obtains the successful detector response. Suppose that Alice prepares the initial state as $|\psi_{a_i}\rangle=\cos\theta|0\rangle+e^{\frac{i2\pi{a_i}}{n}}\sin\theta|1\rangle$. The quantum state after the first round of photon transmission can be viewed as a unitary operation $U(\delta\theta)$ acting on the photon as
\begin{eqnarray}\label{diyilunguangzi}
		|\psi_{a_i}'\rangle&=&U(\delta\theta)|\psi_{a_i}\rangle\nonumber\\
&=&\cos{(\theta+\delta\theta)}|0\rangle+e^{\frac{i2\pi{a_i}}{n}}\sin{(\theta+\delta\theta)}|1\rangle.		
\end{eqnarray}
In the first round of security checking, the practical probability distribution becomes
 \begin{eqnarray}\label{gailv3}
	\begin{aligned}
		\begin{split}
			&P^{'}_1(g=0)=\frac{1}{rQ_{AB}}\sum_{i=1}^{rQ_{AB}}|\cos\theta\cos(\theta+\delta\theta_i)\\
			&+e^{\frac{i2\pi(a_i-w_i)}{n}}\sin\theta\sin(\theta+\delta\theta_i)|^2,\\
			&P^{'}_1(g=1)=1-P^{'}_1(g=0).
		\end{split}
	\end{aligned}
\end{eqnarray}

After two rounds of photon transmission, the error caused by the noise can be viewed as performing the unitary operation $U(2\delta\theta)$ on the transmitted photons. In this way,
the quantum state of the photon in $S_2$ sequence evolves  to
\begin{eqnarray}\label{dierlun}
	\begin{aligned}
		\begin{split}
			|\psi_{b_i}'\rangle=\cos{(\theta+2\delta\theta_i)}|0\rangle+e^{\frac{i2\pi{b_i}}{n}}\sin{(\theta+2\delta\theta_i)}|1\rangle.
		\end{split}
	\end{aligned}
\end{eqnarray}
The practical probability distribution in the second round of security checking becomes
\begin{eqnarray}\label{gailv4}
	\begin{aligned}
		\begin{split}
			&P^{'}_2(g=0)=\frac{1}{rQ_{ABA}}\sum_{i=1}^{rQ_{ABA}}|\cos\theta\cos(\theta+2\delta\theta_i)\\
			&+e^{\frac{i2\pi(b_i-d_i)}{n}}\sin\theta\sin(\theta+2\delta\theta_i)|^2,\\
			&P^{'}_2(g=1)=1-P^{'}_2(g=0).
		\end{split}
	\end{aligned}
\end{eqnarray}

 We use the deviation between the theoretical probability distribution and the practical probability distribution of $g=0$ after the first and second round of photon transmission to calculate the error rates $e_{AB}$ and $e_{ABA}$ caused by the quantum state error, respectively. $e_{AB}$ and $e_{ABA}$ can be calculated as
\begin{eqnarray}\label{liangluncuowu-1}
	\begin{aligned}
		\begin{split}
			e_{AB}&=Q_{AB}[P_1(g=0)-P^{'}_1(g=0)]\\
&=|\frac{\sum_{i=1}^{\eta_t\eta_c\eta_m\eta_dr}\cos({\frac{2\pi{(a_i-w_i)}}{n}})[1-\sin(2\theta+2\delta\theta_i)]}{2r}|,\\
			e_{ABA}&=Q_{ABA}[P_2(g=0)-P^{'}_2(g=0)]\\
&=|\frac{\sum_{i=1}^{\eta^2_t\eta^2_c\eta^2_m\eta_dr}\cos({\frac{2\pi{(b_i-d_i)}}{n}})[1-\sin(2\theta+4\delta\theta_i)]}{2r}|.
		\end{split}
	\end{aligned}
\end{eqnarray}

In step two, we consider the photon loss scenario. During both rounds of security checking, the parties first record the non-click events and calculate the theoretical probability distribution of $g=0$ of the lost photons. If the theoretical probability distribution of $g=0$ is in the scale of $(0, 0.5]$, the parties deterministically define the non-click events as $g=1$. On the contrary, if the theoretical probability distribution of $g=0$ is in the scale of $(0.5, 1]$, the parties define the non-click events as $g=0$. Under this assumption, the simulated value of $C_S$ is symmetrical about $P_1(g=0)=0.5$.
We take the former case as an example. In this case, the no-click events would cause the error rate $e_{AB}'$ and $e_{ABA}'$ as
 \begin{eqnarray}\label{liangluncuowu-1n}
	\begin{aligned}
		\begin{split}
 e_{AB}'=\frac{\sum_{i=1}^{(1-Q_{AB})r}[1+\cos(\frac{2\pi(a_i-w_i)}{n})]}{2r},\\
 e_{ABA}'=\frac{\sum_{i=1}^{(1-Q_{ABA})r}[1+\cos(\frac{2\pi(b_i-d_i)}{n})]}{2r}.
 \end{split}
	\end{aligned}
\end{eqnarray}
In the latter case, we can also calculate the error rate $e_{AB}'$ and $e_{ABA}'$ by replacing $\cos(\frac{2\pi(a_i-w_i)}{n})$ and $\cos(\frac{2\pi(b_i-d_i)}{n})$ in Eq. (\ref{liangluncuowu-1n}) with $-\cos(\frac{2\pi(a_i-w_i)}{n})$ and $-\cos(\frac{2\pi(b_i-d_i)}{n})$, respectively.

Therefore, we can obtain the total error rates $E_{AB}$ and $E_{ABA}$ as
\begin{eqnarray}\label{liangluncuowu-2}
	\begin{aligned}
		\begin{split}
			&E_{AB}=e_{AB}+e_{AB}',\\
			&E_{ABA}=e_{ABA}+e_{ABA}'.
		\end{split}
	\end{aligned}
\end{eqnarray}

 In the numerical simulation process, since the exact details of the lost photons and randomly selected bases are now known, some simplifications are required. The distribution of successfully responded photon measurement results is simplified to an average distribution as
\begin{eqnarray}\label{gailv5}
	\begin{aligned}
		\begin{split}
			&P^{'}_1(g=0)\approx\frac{\eta_t\eta_c\eta_m\eta_d}{r}\sum_{i=1}^{r}|\cos\theta\cos(\theta+\delta\theta_i)\\
			&+e^{\frac{i2\pi(a_i-w_i)}{n}}\sin\theta\sin(\theta+\delta\theta_i)|^2,\\
			&P^{'}_2(g=0)\approx\frac{\eta^2_t\eta^2_c\eta^2_m\eta_d}{r}\sum_{i=1}^{r}|\cos\theta\cos(\theta+2\delta\theta_i)\\
			&+e^{\frac{i2\pi(b_i-d_i)}{n}}\sin\theta\sin(\theta+2\delta\theta_i)|^2.
		\end{split}
	\end{aligned}
\end{eqnarray}
To simplify the calculation of $E_{AB}$ and $E_{ABA}$, we treat the theoretical probability distribution of $g=0$ in two security checking processes to be identical, say,
\begin{eqnarray}\label{gailv6}
	\begin{aligned}
		\begin{split}
			&P_1(g=0)=P_2(g=0).
		\end{split}
	\end{aligned}
\end{eqnarray}
Additionally, in order to determine the threshold for the total detection efficiency, we simplify $Q_{AB}$ and $Q_{ABA}$ as (considering $\eta_d$ is quite close to 1 \cite{detector})
\begin{eqnarray}\label{eta}
	\begin{aligned}
		\begin{split}
			&Q_{AB}=\eta_t\eta_c\eta_m\eta_d\approx\eta,\\
			&Q_{ABA}=\eta^2_t\eta^2_c\eta^2_m\eta_d\approx\eta^2.
		\end{split}
	\end{aligned}
\end{eqnarray}

 In the simulation, we need to assign a value of the channel error $\delta\theta$. In this way, we simplify the channel errors of different photons into a uniform channel error $\delta\theta$ as
 \begin{eqnarray}\label{cuowujianhua}
	\begin{aligned}
		\begin{split}
			&\sum_{i=1}^{r}\cos{(\theta+\delta\theta_i)}|0\rangle+e^{\frac{i2\pi{x_i}}{n}}\sin{(\theta+\delta\theta_i)}|1\rangle\\
			&\rightarrow \sum_{i=1}^{r}\cos{(\theta+\delta\theta)}|0\rangle+e^{\frac{i2\pi{x_i}}{n}}\sin{(\theta+\delta\theta)}|1\rangle.
		\end{split}
	\end{aligned}
\end{eqnarray}

\begin{figure*}[t]
	\centering
	\includegraphics[scale=0.48]{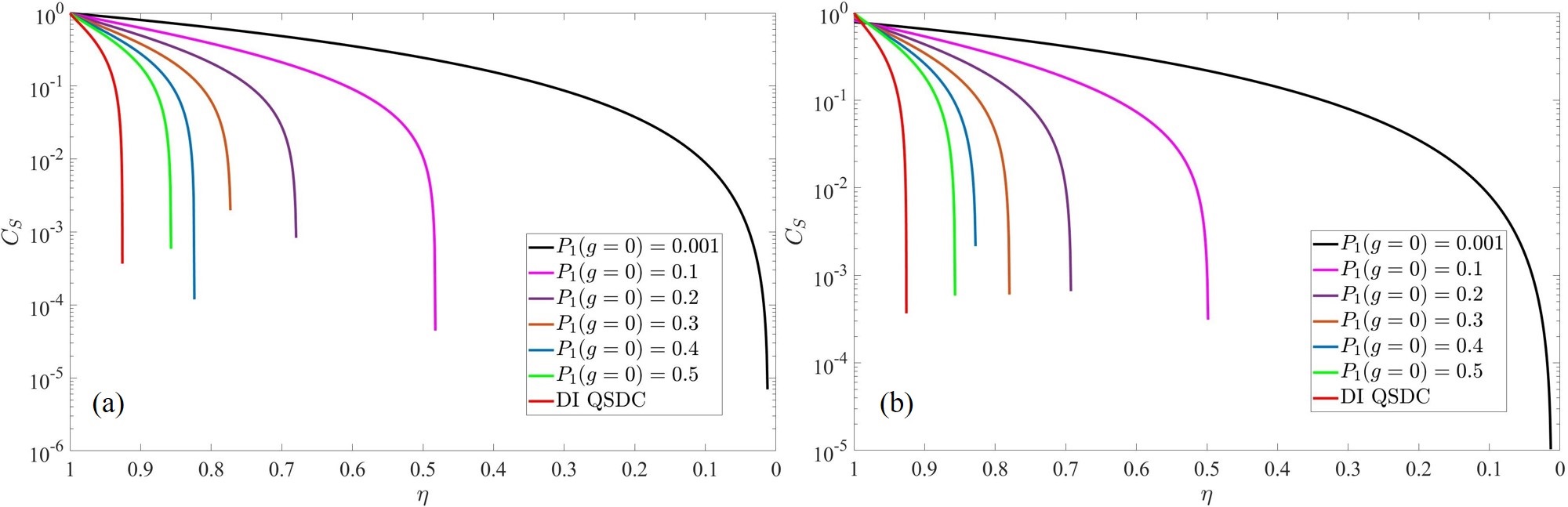}
	\caption{The secrecy message capacity $C_{S}$ of the RDI QSDC protocol and DI QSDC protocol \cite{DIQSDC1} versus the total detection efficiency $\eta$. In RDI QSDC protocol, we control  $\theta=\frac{\pi}{4}$, and adjust $P_1(g=0)=0.001, 0.1, 0.2, 0.3, 0.4, 0.5$, respectively. The noise operator $\delta\theta$ is set to be  $\frac{\pi}{400}$ (a) and $\frac{\pi}{40}$ (b), respectively. In DI QSDC, the fidelity of the target entangled is set to be 1.}
	\label{Fig3}
\end{figure*}

In Fig. \ref{Fig3}, we control $\theta=\frac{\pi}{4}$ and set the noise operator $\delta\theta=\frac{\pi}{400}$ (relatively low noise) and $\frac{\pi}{40}$ (relatively high noise), and simulate $C_S$ of the RDI QSDC protocol altered with the total detection efficiency $\eta$ under different values of $P_1(g=0)$. We also provide the $C_S$ of DI QSDC protocol with the fidelity $F=1$. For the RDI QSDC protocol, $C_{S}$ and the threshold of $\eta$ is largely influenced by the theoretical probability distribution $P_1(g=0)$. The lower value of $P_1(g=0)$ leads to the higher $C_{S}$ and lower threshold of $\eta$.
In detail, with $\delta\theta=\frac{\pi}{400}$, the threshold of $\eta$ is 0.0120, 0.4825, 0.7728, 0.8238, 0.8569 corresponding to $P_1(g=0)$ being 0.001, 0.1, 0.2, 0.3, 0.4, 0.5, respectively. If we set $\delta\theta=0$, the corresponding threshold of $\eta$ further reduces to 0.0115, 0.4823, 0.6790, 0.7718, 0.8238, 0.8568, respectively (not shown in Fig. \ref{Fig2}), which is far lower than that of DI QSDC (about 0.926).
 Here, we set $\eta_{c}=0.95$, $\eta_{m}=\eta_{d}=1$ \cite{detector}, the maximal secure communication distance of our RDI QSDC protocol reaches 95.81 km and 14.72 km with $P_1(g=0)=0.001, 0.1$, respectively. When $\delta\theta=\frac{\pi}{40}$, the threshold of $\eta$ slightly  increases to 0.0130, 0.4985, 0.6927, 0.7798, 0.8278, 0.8569 with $P_1(g=0)=0.001, 0.1, 0.2, 0.3, 0.4, 0.5$, respectively. It can be found that with the growth of $P_1(g=0)$, the influence of $\delta\theta$ on the threshold of $\eta$ reduces. For example, in the case of  $P_1(g=0)=0.001$, the threshold of $\eta$ increases from 0.01201 to 0.01301 (about 8.33\% growth) with $\delta\theta$ increasing from  $\frac{\pi}{400}$ to $\frac{\pi}{40}$, but with $P_1(g=0)=0.4$, the threshold of $\eta$ increases from 0.8238 to 0.8278 (about 0.49\%). It indicates the higher values of $P_1(g=0)$ leads to the stronger noise robustness.

\begin{figure}[htbp]
	\centering
	\includegraphics[scale=0.16]{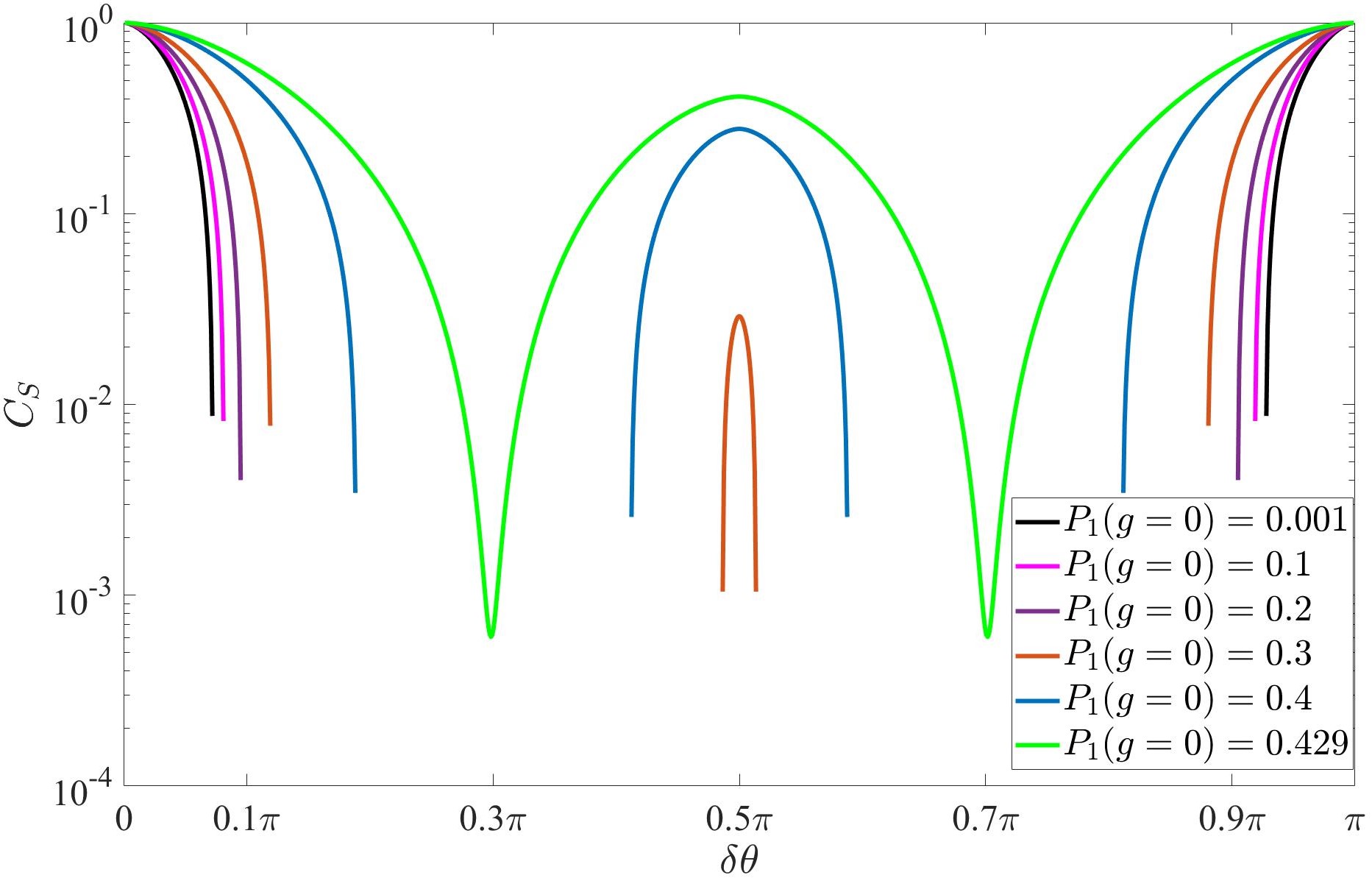}
	\caption{The secrecy message capacity $C_{S}$ altered with $\delta\theta$ under $P_1(b=0)=0.001, 0.1, 0.2, 0.3, 0.4, 0.429$, respectively.}
	\label{Fig4}
\end{figure}

 For further researching the noise robustness of our RDI QSDC protocol, we assume no photon loss and calculate $C_S$ versus $\delta\theta$ under $P_1(g=0)=0.001, 0.1, 0.2, 0.3, 0.4, 0.429$, respectively.
  From Eq. (\ref{liangluncuowu-1}),  the periods of $\delta\theta$ in $e_{AB}$ and $e_{ABA}$ are $\pi$ and $0.5\pi$, respectively, so that we research the influence of  $\delta\theta$ in the range of $[0,\pi]$. It can be found that $C_{S}$ is axisymmetric about $\delta\theta=0.5\pi$. With the growth of $P_1(g=0)$, the range of tolerable $\delta\theta$ becomes larger, indicating that the protocol has stronger noise robustness. In particular, when $P_1(g=0)=0.429$, which is quite close to 0.5, $C_{S}$ has the positive value in the whole range of $[0,\pi]$. That is to say, our protocol can resist the noise in the whole range of $[0,\pi)$. It is interesting that $C_{S}$ obtains a relatively high value at $\delta\theta=0.5\pi$ with high value of $P_1(g=0)$ (0.429, 0.4, 0.3). The reason is that in above three cases, $e_{ABA}$ obtains the minimal value at $\delta\theta=0.5\pi$ and $e_{AB}$ is at a relatively low level, which can effectively increase $C_{S}$. However, under the case of low $P_1(g=0)$ ($0.2, 0.1, 0.001$), $e_{AB}$ is at a high level with $\delta\theta=0.5\pi$. In this way, although $e_{ABA}$ obtains the minimal value, $C_{S}$ can not obtain the positive value. When $\delta\theta$ is in the scale of $(0,0.3\pi)$, the threshold of $\delta\theta$ is 0.2547, 0.2988, 0.3742, 0.5912 corresponding to $P_1(g=0)=0.1, 0.2, 0.3, 0.4$, respectively. We define the fidelity $F$ as the inner product of the degraded quantum state after two rounds of photon transmission and the initial quantum state. The threshold of $F$ with $P_1(g=0)=0.1, 0.2, 0.3, 0.4$ can be calculated as 0.9365, 0.9133, 0.8664, 0.6894, respectively.

 From above discussion, low total detection efficiency threshold requires $P_1(g=0)$ to deviate from 0.5 but high noise robustness requires it to be close to 0.5. There is a trade-off between the total detection efficiency threshold and noise robustness. In practical applications, we have to consider both two factors and choose suitable theoretical distribution of $P_1(g=0)$.

\section{Discussion and Conclusion}\label{Section5}
Our paper proposes a single-photon based RDI QSDC protocol, which treats the whole receiving devices in both parties as ``black-boxes''. In our protocol, we require $\theta=\frac{\pi}{4}$, $n\geq3$ and $n\neq4$. When $n=2$, Alice can prepare photons with only X basis $\{\frac{1}{\sqrt{2}}(|0\rangle\pm|1\rangle)\}$. The QSDC protocol cannot ensure its security with only one basis. When $n=4$, Alice can prepare photons with two bases, say, X basis and Y basis $\{\frac{1}{\sqrt{2}}(|0\rangle\pm i|1\rangle)\}$. In this case, the theoretical value of $P_1(g=0)$ will deterministically be 0.50. From Sec. \ref{Section3}, in this case, the protocol can not resist the blinding attack. In other cases, we can ensure Alice and Bob have more than two bases, which can ensure the security of the protocol. Our protocol is more sensitive to the blinding combined with fake-state attack with higher value of $n$. The specific value of $n$ can be determined by establishing an evaluation model based on the resources consumed, the accuracy of the probability distribution and the secure message capacity. We can use some methods like simulated annealing \cite{tuihuo1,tuihuo2,tuihuo3} to seek an optimal value for $n$, which makes the RDI QSDC protocol consume fewer single photon resources and maintain relatively high accuracy. On the other hand, in the paper, we consider the on-demand single photon source. If we consider the practical imperfect single photon source, the photon number splitting (PNS) attack from the multi-photon event and the side-channel attack may cause the message leakage. The adoption of the decoy-state method \cite{PNS} and passive method \cite{ying,passive1,passive2} into the RDI QSDC protocol is promising to reduce the message leakage from multi-photon events. We will research on the above topics in our future works.

It is interesting to compare our RDI QSDC protocol with existing DI QSDC \cite{DIQSDC1,DIQSDC2,DIQSDC3} and MDI QSDC \cite{MDIQSDC1,MDIQSDC2} protocols. First, DI QSDC and MDI QSDC protocols  \cite{DIQSDC1,DIQSDC2,MDIQSDC1,MDIQSDC2} require the high-quality entanglement source and BSM. Current entanglement sources have relatively low efficiency in the range of $10^{-5}-10^{-3}$ \cite{Hu}. BSM with linear optical elements has the success probability of 50\%. Current low-efficient entanglement source and BSM would largely reduce the practical communication efficiency of DI QSDC and MDI QSDC. Our RDI QSDC protocol only requires the single-photon source. Current practical single-photon sources have already allowed for nearly on-demand \cite{demand},
highly-efficient \cite{highlyefficient} extraction of single photons, can maintain the purity and indistinguishability of the emitted photons above 99\% \cite{99,ondemand}. In 2020, the experimental implementation of free-space QSDC based on single-photon source with the repetition rate of 16 MHz was realized \cite{QSDCe3}. Meanwhile, the RDI QSDC only requires SPM, which is easy to operate and has high efficiency (close to 100\$). Second, DI QSDC \cite{DIQSDC1,DIQSDC2} depends on extremely high-quality entanglement or hyperentanglement distribution to maintain the nonlocal correlation of the photons, so that it requires quite high global detection efficiency (about 0.9258) and have quite low noise tolerance (the fidelity threshold of the target state is above 0.95). The maximal secure communication distances of DI QSDC is quite short. From the numerical simulation results in Sec. \ref{Section4}, by selecting a suitable value of $P_{1}(g=0)$, our RDI QSDC protocol can achieve relatively high maximal secure communication distance and low fidelity threshold.

\begin{figure}[htbp]
	\centering
	\includegraphics[scale=0.16]{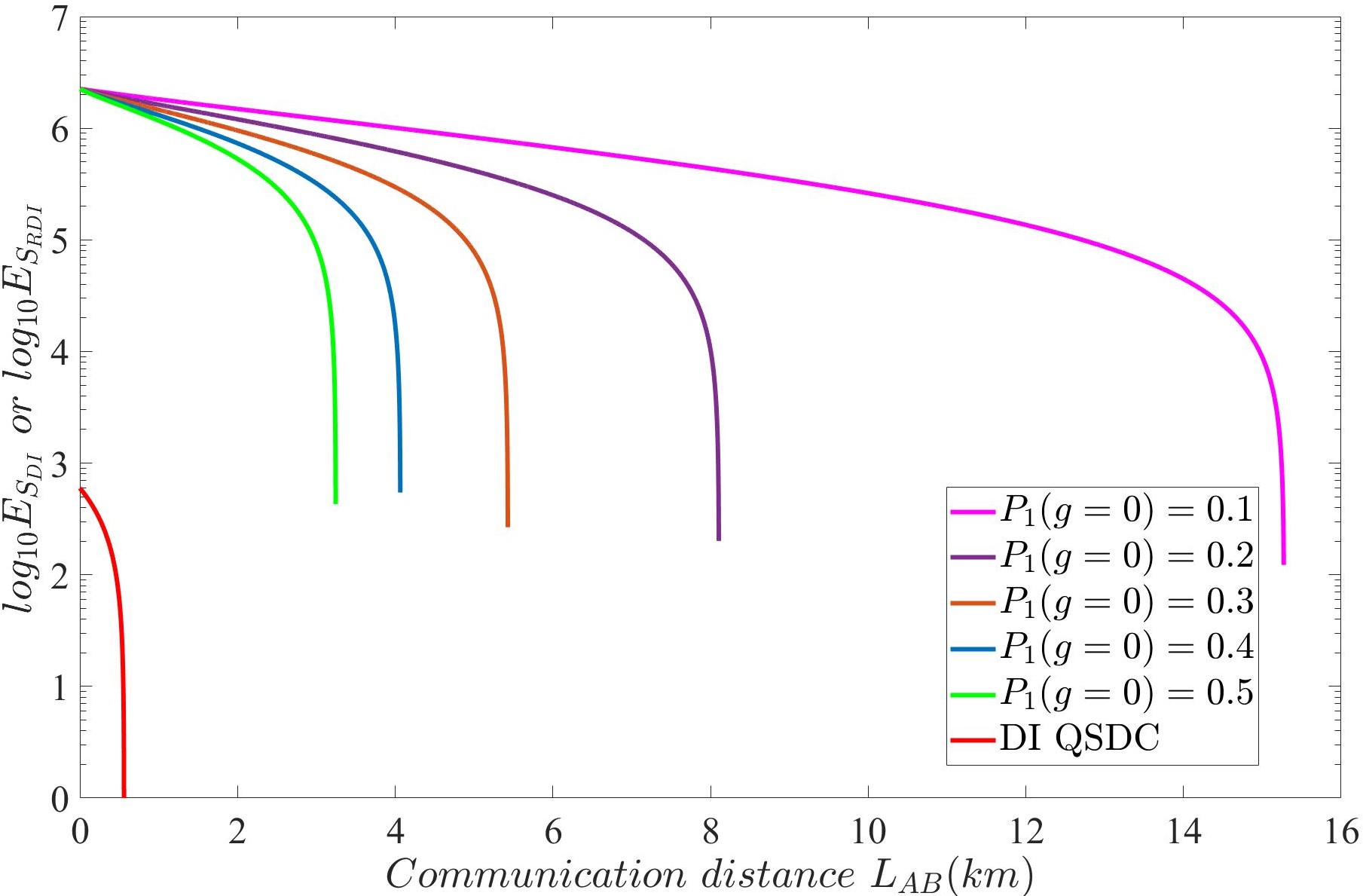}
	\caption{The practical communication efficiency $E_{s}$ of the RDI QSDC protocol and DI QSDC protocol \cite{DIQSDC1} versus the communication distance $L$. Here, we control $\eta_{c}=0.95$, $\eta_{m}=\eta_{d}=1$, $p_{e}=10^{-3}$, and $p_{s}=1$. In the RDI QSDC protocol, we set $P_1(b=0)=0.1, 0.2, 0.3, 0.4, 0.5$, respectively.}
	\label{Fig5}
\end{figure}

Here, we define the practical communication efficiency $E_{s}$ as the amount of transmitted correct and
secure qubits per second. Here, we suppose the repetition rate of the signal source as $R_{rep}$. The generation efficiencies of the entanglement source and single-photon source are defined as $p_{e}$ and $p_{s}$, respectively. In DI (RDI) QSDC, suppose that after each round of photon transmission, we choose half number of entangled
photon pairs (single photons) to perform the DI (RDI) security checking, so that
only $\frac{1}{4}$ of the amount of entangled photon pairs (single photons) can be
used to transmit messages. In this way, we can calculate
$E_{s}$ of the DI QSDC protocol \cite{DIQSDC1} and the current RDI QSDC protocol  as
\begin{eqnarray}\label{E}
			E_{s_{DI}}&=&\frac{1}{4}R_{rep}p_{e}C_{S_{DI}}, \nonumber\\
			E_{s_{RDI}}&=&\frac{1}{4}R_{rep}p_{s}C_{S_{RDI}}.
\end{eqnarray}
Here, we set $R_{rep}=10$ MHz, $\eta_{c}=0.95$, $\eta_{m}=\eta_{d}=1$, $p_{e}=10^{-3}$, and $p_{s}=1$. $E_{s_{RDI}}$ and $E_{s_{DI}}$ altered with the communication distance $L$ are shown in Fig. \ref{Fig5}. In the RDI QSDC protocol, we adjust $P_1(b=0)=0.1, 0.2, 0.3, 0.4, 0.5$, respectively. Compared with DI QSDC, RDI QSDC has obvious advantages in both maximal secure communication distance and practical communication efficiency.  For example, DI QSDC's maximal secure communication distance is only 0.561 km, and RDI QSDC's maximal secure communication distance with $P_1(b=0)=0.1$ achieves 14.72 km, which is about 26 times of that in DI QSDC. With the communication distance of 0.5 km, $E_{s_{RDI}}$ with $P_1(b=0)=0.1$ is about 3415 times of $E_{s_{DI}}$.

QM plays a key role in QSDC, for the parties should first ensure the security of photon transmission and then perform the encoding and decoding operations. Meanwhile, QM is also widely used in some QKD protocols to enhance QKD's communication distance and key generation rate \cite{qmqkd2,qmqkd1,qmqkd5,qmqkd6}. For example, some MDI QKD protocols adopt the QM to improve the photon synchronization, thereby increasing the success probability of BSM \cite{qmqkd2,qmqkd1,qmqkd5}. In this paper, we adopt the all-optical, polarization insensitive storage loop as the QM. The storage loop was first designed in the experimental preparation of four-photon and six-photon
GHZ states \cite{QM}. In Ref. \cite{QM}, the storage loop stores the photons with a
central wavelength of 1550 nm and a bandwidth of 0.52
THz. It achieves the storage efficiency
of 91\% and a lifetime of 131 ns, corresponding to around 11 round trips. Comparing with the common solid QMs, such as the atomic QMs (memory lifetime is in the $\mu$s scale) \cite{cold1,cold2,cold3,2019wang,2021bao,2022buser}, the all-optical storage loop QM has shorter memory lifetime. However, it is feasible under current experimental condition and can operate at any wavelength in principle with only the
minor adaptions. In contrast, the atomic QMs only can store the photons in a narrow wavelength scale based on the atomic
level structure of the underlying material system. In this way, the all-optical storage loop QM is more flexible and has wide application in future quantum communication field.

In summary, QSDC enables the message sender to directly send secret messages to the receiver through quantum channels without keys. The practical imperfect experimental devices bring security loopholes to QSDC.  Existing DI and MDI QSDC protocols can enhance QSDC's security in theory. However, DI QSDC requires extremely high global detection efficiency and has quite low secure communication distance.  Meanwhile, the requirements of high-quality entanglement source largely limit DI QSDC's and MDI QSDC's practical communication efficiency.
In the paper, we propose a single-photon-based receiver-device-independent (RDI) QSDC protocol. The RDI QSDC protocol only relies on the trusted single-photon source, which is nearly on-demand and high-efficient under current technology, and treats all the receiving devices in both communication parties as ``black-boxes''. The parties ensure the message security only from the observed statistics, say, the deviation between the theoretical measurement result distribution and the practical measurement result distribution. The RDI QSDC is immune to all possible attacks on the receiving devices and provides the same security level as MDI QSDC. We research on RDI QSDC's performances and develop a numerical method to simulate its secrecy message capacity in practical noisy communication situations.
 Compared with DI QSDC and MDI QSDC, the RDI QSDC has some advantages. First,
it only uses the single-photon source and single photon measurement, which are high-efficient and easy to operate under current experimental conditions.  In particular, it adopts the all-optical storage loop as the QM, so that the whole protocol is feasible with current experimental technology. Second, it has higher photon loss robustness and noise tolerance than DI QSDC, which enables RDI QSDC to achieve higher secure communication distance. With $P_1(b=0)=0.1$, RDI QSDC's maximal secure communication distance achieves 14.72 km, about 26 times of that in DI QSDC. With the communication distance of 0.5 km, RDI QSDC's practical communication efficiency  with $P_1(b=0)=0.1$ is about 3415 times of that in DI QSDC.
Based on above features, our RDI QSDC protocol makes it possible to achieve highly secure and feasible QSDC with current technology.

\section*{Acknowledgement}\label{Acknowledgement}
This work was supported by the National Natural Science Foundation of China under Grant Nos. 12175106 and 92365110, and the Postgraduate Research $\&$ Practice Innovation Program of Jiangsu Province under Grant NO. KYCX24-1127.

\appendix
\setcounter{equation}{0}
\setcounter{subsection}{0}
\renewcommand{\theequation}{A.\arabic{equation}}

\section{All-optical, polarization insensitive storage loop QM} \label{appendixA}
 QM is the key element of QSDC.  In our RDI QSDC protocol, we adopt the all-optical, polarization insensitive storage loop as the QM.  In Fig. \ref{Fig2}(b), we provide the schematic principle  of this all-optical storage loop QM. This storage loop only uses the linear optical polarization beam splitter (PBS), quarter-wave-plate (QWP), common electro-optic modulator (EOM) and field programmable gate array
(FPGA) to achieve the functions of storage and readout of single photons. The on-off of the EOM is controlled by the FPGA. When the EOM is turned on, the polarization of the passing photon will be flipped. When EOM
is turned off, the polarization of the passing photon will
not change. The function of the PBS is to totally transmit the photon in $|H\rangle$ and totally reflect the photon in $|V\rangle$.  The function of the EOM can be described as
 \begin{eqnarray}\label{EOM}
	\begin{aligned}
		\begin{split}
			|H\rangle&\xrightarrow{EOM(ON)}&|V\rangle,\qquad
|H\rangle\xrightarrow{EOM(OFF)}&|H\rangle,\\
|V\rangle&\xrightarrow{EOM(ON)}&|H\rangle,\qquad
|V\rangle\xrightarrow{EOM(OFF)}&|V\rangle.
		\end{split}
	\end{aligned}
\end{eqnarray}

 The specific process of QM is as follows. When a photon first passes through the PBS and enters the storage loop, the EOM is in the OFF state. After the photon passing through EOM, EOM switches to the ON state, allowing the photon to continue to circulate within the storage loop. When the party hopes to extract the photon, he switches EOM to the OFF state, allowing the photon to exit the QM. The basic principle of the QM can be described as
 \begin{widetext}
 \begin{eqnarray}\label{QM}
	\begin{aligned}
		\begin{split}
			&|H\rangle_{a}^{in}\xrightarrow{PBS}|H\rangle_{c}\xrightarrow{EOM(OFF)}|H\rangle_{d}\xrightarrow{PBS}|H\rangle_{b}\xrightarrow{twice\  QWP}|V\rangle_{b}\xrightarrow{PBS}|V\rangle_{c}\xrightarrow{EOM(ON)}|H\rangle_{d}\\
			&\xrightarrow{loop\ with\ EOM\ turn\ on}\cdots\xrightarrow{exit\ loop}|V\rangle_{b}\xrightarrow{PBS}|V\rangle_{c}\xrightarrow{EOM(OFF)}|V\rangle_{d}\xrightarrow{PBS}|V\rangle_{a}^{out},\\
			&|V\rangle_{a}^{in}\xrightarrow{PBS}|V\rangle_{d}\xrightarrow{EOM(OFF)}|V\rangle_{b}\xrightarrow{PBS}|V\rangle_{b}\xrightarrow{twice\  QWP}|H\rangle_{b}\xrightarrow{PBS}|H\rangle_{d}\xrightarrow{EOM(ON)}|V\rangle_{b}\\
			&\xrightarrow{loop\ with\ EOM\ turn\ on}\cdots\xrightarrow{exit\ loop}|H\rangle_{b}\xrightarrow{PBS}|H\rangle_{d}\xrightarrow{EOM(OFF)}|H\rangle_{b}\xrightarrow{PBS}|H\rangle_{a}^{out}.
		\end{split}
	\end{aligned}
\end{eqnarray}
\end{widetext}
It can be found that after the photon exiting the QM, its polarization will be flipped.  After the photon is read out, we can pass the photon through an half-wave plate (HWP)  to recover its polarization feature.

\end{document}